\documentclass[conference]{IEEEtran}

\usepackage{graphicx}
\usepackage{subfigure}
\usepackage{amsmath}
\usepackage{amsfonts}
\usepackage{amssymb}
\usepackage{latexsym}
\usepackage{moreverb}
\usepackage{booktabs}
\usepackage{times}
\usepackage{indentfirst}
\usepackage{algorithmic}
\usepackage{algorithm}
\usepackage[table]{xcolor}  
\usepackage{multirow}       
\usepackage{array}
\usepackage{picins}
\usepackage{wrapfig}
\usepackage{fancyhdr}
\graphicspath{{figures/}}

\newcommand{\Fig}[1]{Fig.~\ref{#1}}

\fancypagestyle{IEEE_Green_open_access_footer}{%
  \fancyhf{}
  
  \fancyfoot[C]{\footnotesize \copyright2017 IEEE. Personal use of this material is permitted. Permission from IEEE must be obtained for all other uses, in any current or future media, including reprinting/republishing this material for advertising or promotional purposes, creating new collective works, for resale or redistribution to servers or lists, or reuse of any copyrighted component of this work in other works. DOI: 10.1109/ICC.2017.7997426}
}

\begin{document}

\title{Together or Alone: Detecting Group Mobility \\ with Wireless Fingerprints}

\author{G\"urkan~Solmaz and Fang-Jing~Wu  \\
NEC Laboratories Europe, Heidelberg, Germany\\
\{gurkan.solmaz,fang-jing.wu\}@neclab.eu}

\maketitle
\thispagestyle{IEEE_Green_open_access_footer}

\begin{abstract}
This paper proposes a novel approach for detecting groups of people that walk ``together'' (group mobility) as well as the people who walk ``alone'' (individual movements) using wireless signals. We exploit multiple wireless sniffers to pervasively collect human mobility data from people with mobile devices and identify similarities and the group mobility based on the wireless fingerprints. We propose a method which initially converts the wireless packets collected by the sniffers into people's wireless fingerprints. The method then determines group mobility by finding the statuses of people at certain times (dynamic/static) and  the space correlation of dynamic people. To evaluate the feasibility of our approach, we conduct real world experiments by collecting data from 10 participants carrying Bluetooth Low Energy (BLE) beacons in an office environment for a two-week period. The proposed approach captures space correlation with 95\% and group mobility with 79\% accuracies on average. With the proposed approach we successfully 1) detect the groups and individual movements and 2) generate social networks based on the group mobility characteristics.
\end{abstract}

\begin{keywords}
crowd mobility, human mobility, internet of things, social networks.
\end{keywords}

\IEEEpeerreviewmaketitle

\section{Introduction}
\label{Introduction}

Human mobility analytics has attracted attention for many promising service domains such as public transport~\cite{Zaslavsky13}, public safety~\cite{Kopaczewski15,Pretorius15}, and smart cities. As a result of the advancements of the emerging technologies in Internet of Things (IoT) and communications, human mobility information can be pervasively collected through mobile devices, RFIDs, and sensors for further human activity inference~\cite{Wu16}.

\begin{figure}
\centering
\includegraphics[width=0.45\textwidth]{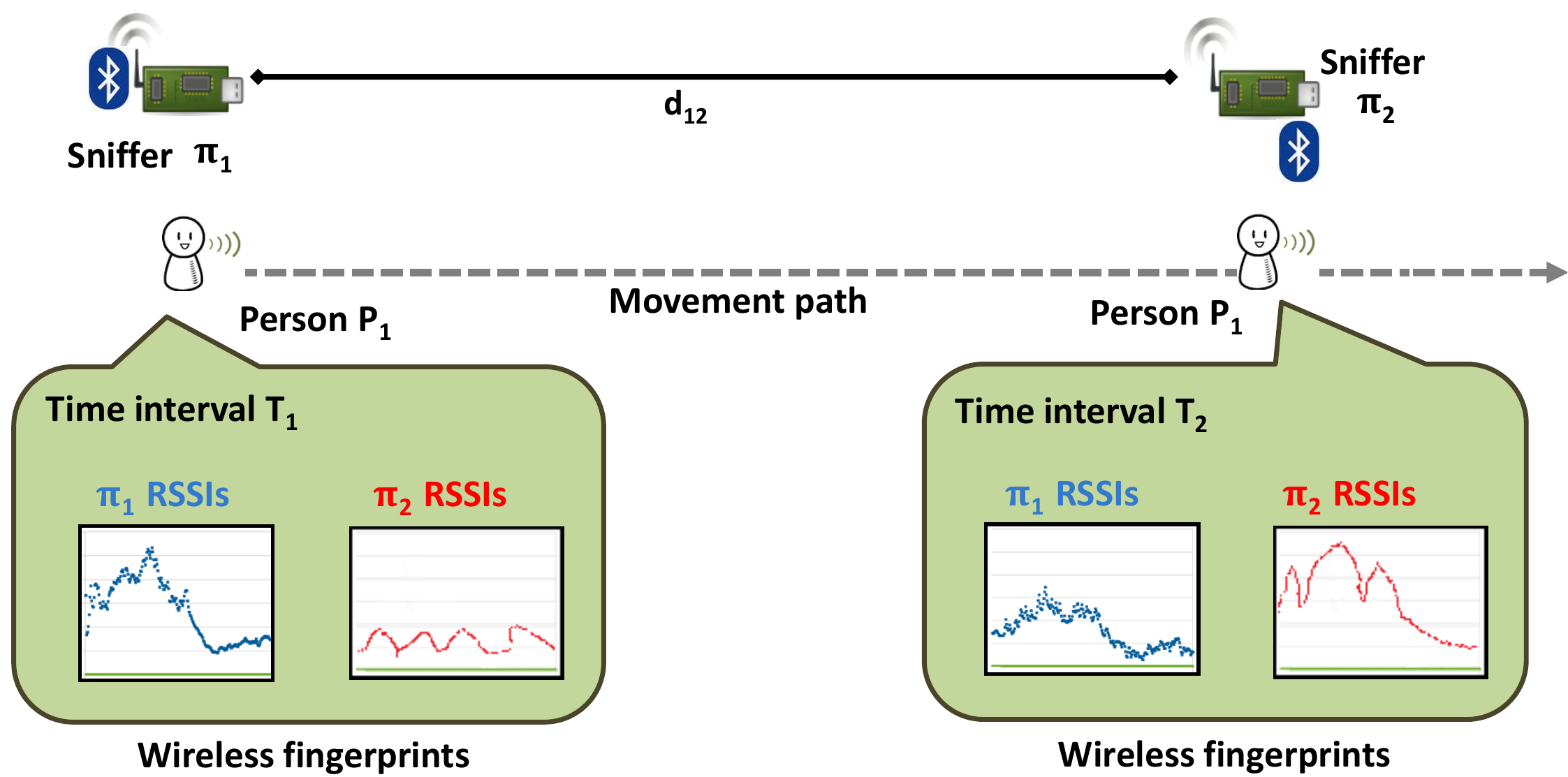}
\caption{Using multiple sniffers for human movement detection.}
\label{Fig:IntroFigure}
\end{figure}

Can we know if there exist groups of people who are walking \emph{together} or people who are always walking \emph{alone}? Comparing to individual activity inference, group mobility emphasizes crowd behaviour~\cite{Bellomo12,Helbing11} more from social perspectives and opens up new opportunities for enhancing human well-being. For example, if we can detect students walking alone or together in a campus, it will be very helpful for understanding social isolation at an earlier stage. Another example is understanding the characteristics of people in certain areas. For instance, if we know that the people who visit a tourist attraction at different times mostly consist of families, couples, or singles, we can do planning of new events based on this knowledge. Thus, to answer these questions and possible others, this paper proposes a novel approach for detecting groups of people that walk together (group mobility) and the people who walk alone (individual movements) using wireless signals. Wireless sniffers are deployed in targeted areas for collecting wireless signals (e.g., Wi-Fi/Bluetooth signals) from mobile devices carried by people. We transform the collected wireless packets to wireless fingerprints of people's movement and determine if these fingerprints are similar to each other in terms of their mobility statuses (i.e., static/dynamic) and their correlation of space transition. Figure~\ref{Fig:IntroFigure} shows an example of wireless signals received during a person's movement which can be used for fingerprints. When he/she moves from one area to another that are covered by different sniffers, the wireless signal strengths vary. Our key idea is to detect if there exist a group of people who have similar wireless fingerprints during their movements. This enables detecting movement groups for surveillance or profiling in certain areas. Moreover, for certain scenarios such as university campuses, we can further identify people's long-term group behaviours and social interactions in crowds.

The proposed approach consists of the phases of sampling and aggregation, wireless fingerprinting, movement detection, space correlation, and group mobility decision. The aim of the approach is to capture groups of people that move together or alone. We evaluate our approach with a real-world experiment by collecting data from 10 participants carrying Bluetooth Low Energy (BLE) beacons in an office environment for two weeks. The results of the real-world experiment show that the proposed approach captures similarity (space correlation) with 95\% and group movements with 79\% accuracies on average. With the proposed approach, we successfully detect the group movements and generate social networks based on the group mobility characteristics.

While mostly computer vision-based studies~\cite{Shao14,Chen13} aimed to tackle group mobility detection using cameras, our approach has the following unique features benefited by leveraging wireless signals captured from people's devices. First, our approach does not require any priori knowledge in the sense that no training stage is required. Second, since our approach does not rely on any localization technology, it is flexible to indoor and outdoor environments and to various conditions such as darkness, blind spots, and behind the walls. Third, compared to the camera-based approaches, our approach has much lower computational cost because of the smaller sizes of collected wireless data. In our experiments, computation of the data collected from 10 people for one day takes less than 2.5 min with a personal computer.

Recently, many research work has paid attention to develop computational methods and understand human behaviour using smartphones and wearable devices. The work in~\cite{Wang2015_SmartGPA} is based on collecting data from students' smartphones to understand how behavioral differences and environmental factors affect students' learning during college. Extracting feature patterns of human walking behaviour based on smartphones data is presented in~\cite{Huang_2016_CyclicFeatures}. In~\cite{Luo_2013_SocialWeaver}, multiple smartphones collaborate to find out the conversation groups nearby. The system in~\cite{Lu_2011:_SpeakerSense} exploits continuous audio sensing to identify the person you are talking with in order to avoid the awkward situation of forgetting his/her name. A group-based navigation system is designed in~\cite{Higuchi_2012_CrowdNavi} to help users find a particular person in a social venue. Compared to these approaches, our approach is more flexible and does not require a mobile application to be pre-installed in users' smartphones.

\section{Group Mobility Detection}
\label{sec:Method}

\subsection{Mobility Sensing System}

\begin{figure}
\centering
\includegraphics[width=0.5\textwidth]{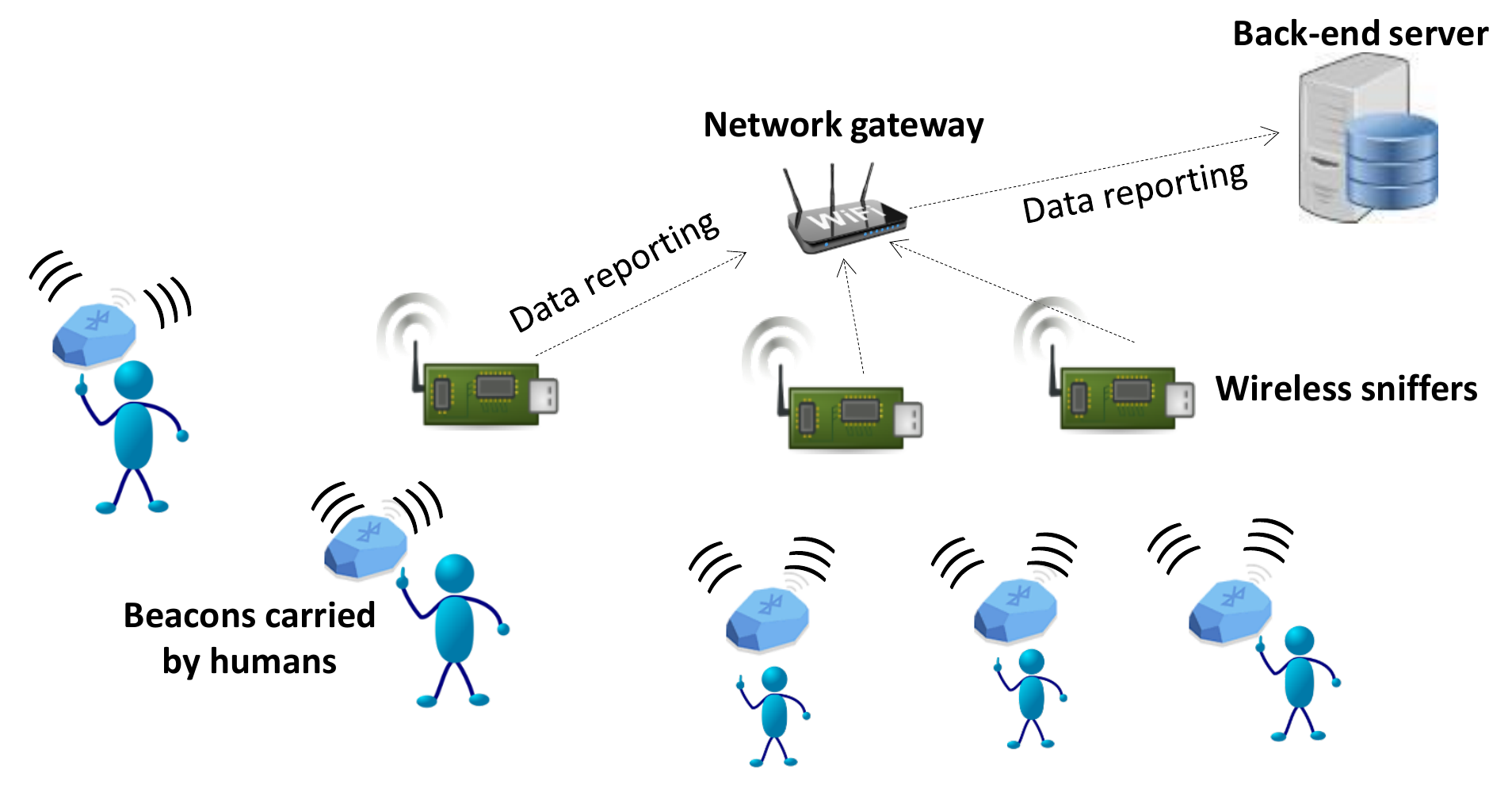}
\caption{System design.}
\label{Fig:SystemDesign}
\end{figure}

The mobility sensing system consists of four components: \emph{beacons}, \emph{wireless sniffers}, a \emph{network gateway}, and a \emph{back-end server}, as shown in \Fig{Fig:SystemDesign}. Each user carries a Bluetooth Low Energy (BLE) beacon which periodically broadcasts advertising packets. Wireless sniffers are deployed randomly in the targeted sensing environment. Wireless sniffers have the capability to capture BLE advertising packets. Each BLE advertising packet contains a unique ID and Received Signal Strength Indicator (RSSI). Wireless sniffers report these captured packets to the network gateway and then to the back-end server for further human mobility data analytics. While we implement a beacon based sniffing system as a prototype for opt-in data collection~\footnote{For experimental purpose, we collect data only from a specific set of BLE beacons carried by voluntary  participants.}, our group mobility detection approach is applicable to the Wi-Fi-based solutions using Wi-Fi sniffers and wireless mobile devices of people (e.g., smartphones).

\begin{figure}
\centering
\includegraphics[width=0.45\textwidth]{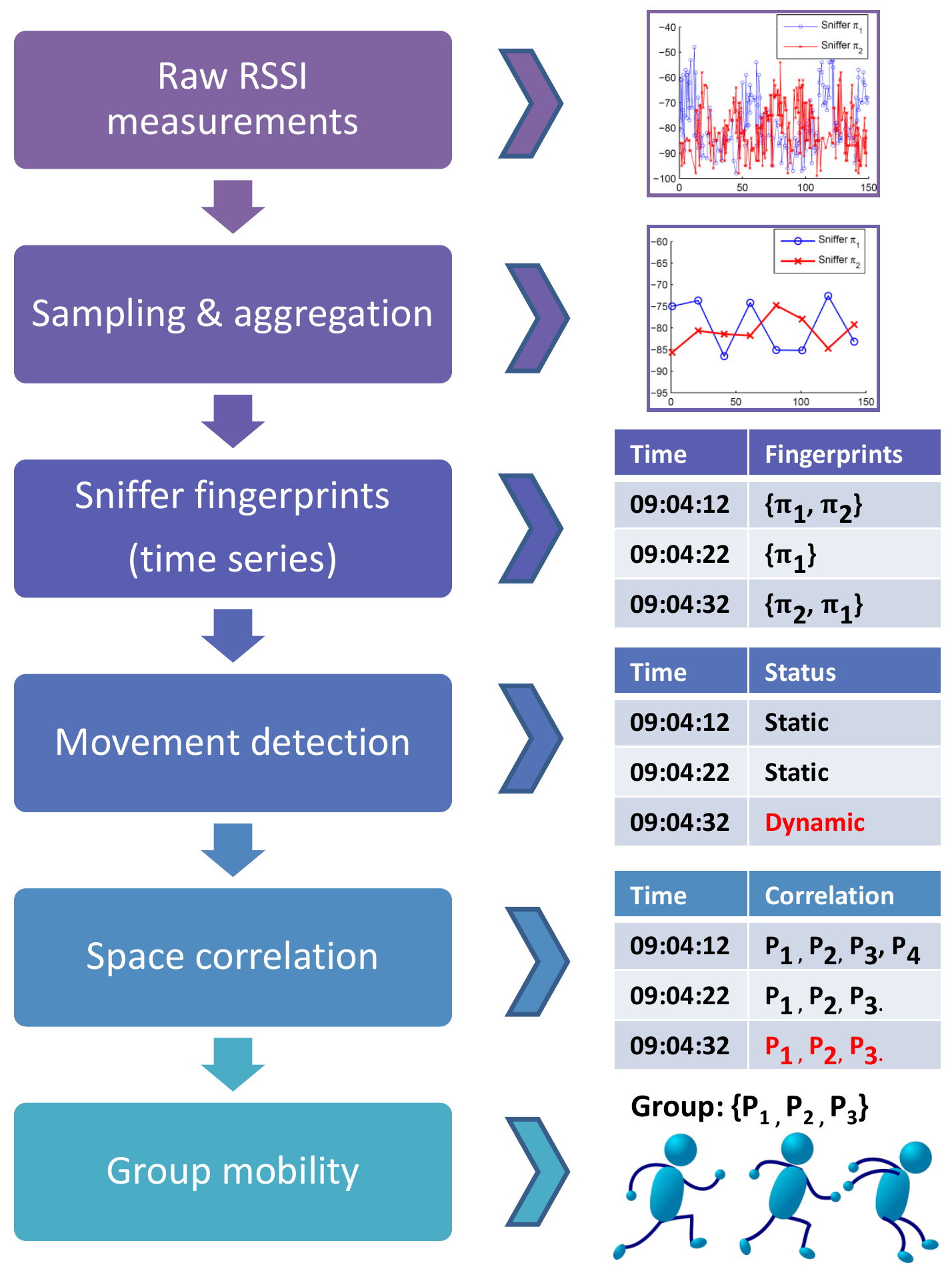}
\caption{The overview of the group mobility detection method.}
\label{Fig:Method}
\end{figure}

Figure~\ref{Fig:Method} illustrates an overview of our algorithm which includes the processes of mobility data preprocessing (sampling and aggregation), wireless fingerprinting (sniffer fingerprints), movement detection (dynamic/static statuses), and space correlation. The mobility data preprocessing is to eliminate noise from raw data. The wireless fingerprinting is to represent the collected data with a list of sniffers sorted by RSSIs. The movement detection is to reduce the search space by detecting if a person is static or dynamic so that the movement detection is done for the smaller set of data (data collected during the person's movement). The space correlation is to extract the levels of the mobility dependency between people for detecting group and individual mobility. Notice that the proposed method can be applied to arbitrary deployment in the sense that one can deploy any number of sniffers randomly or strategically. Below, we describe each of these processes in detail.

\subsection{Mobility Data Preprocessing}

\begin{figure*}[!t]
\centering
\begin{tabular}{ccc}
\includegraphics[width=0.33\linewidth]{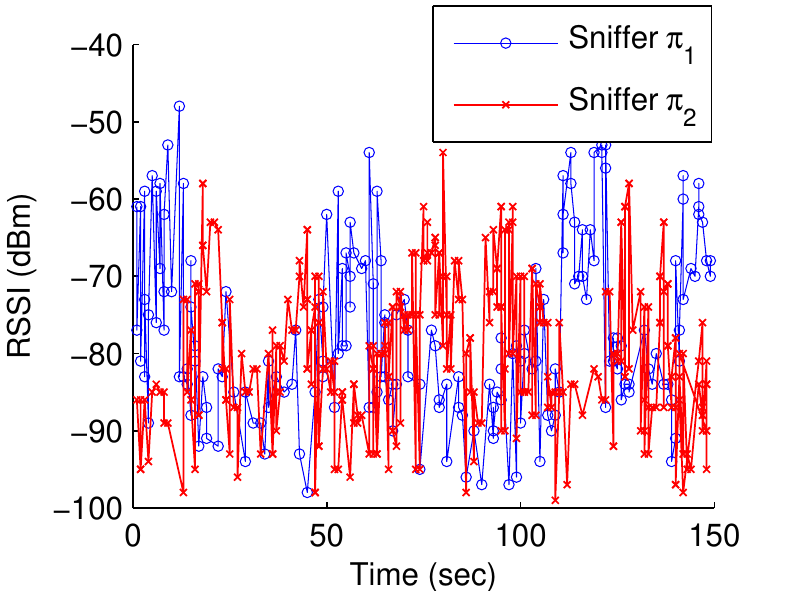} &
\includegraphics[width=0.33\linewidth]{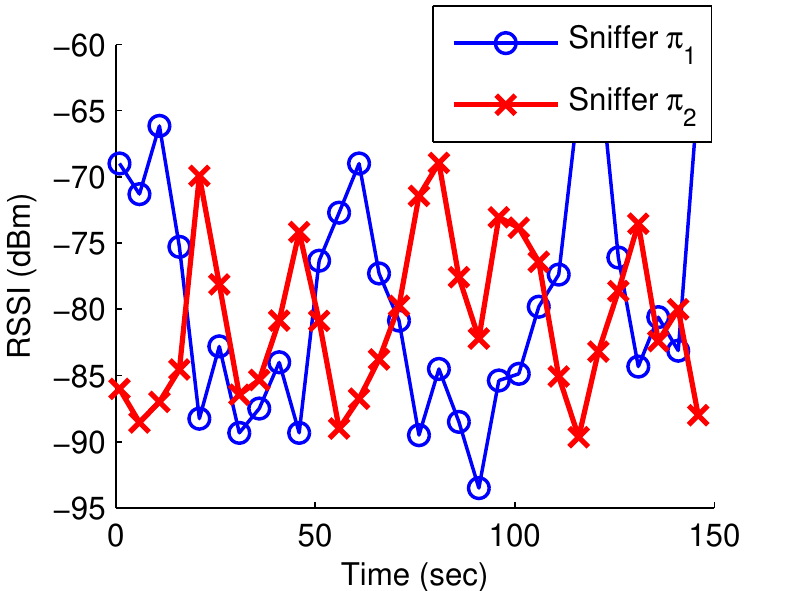} &
\includegraphics[width=0.33\linewidth]{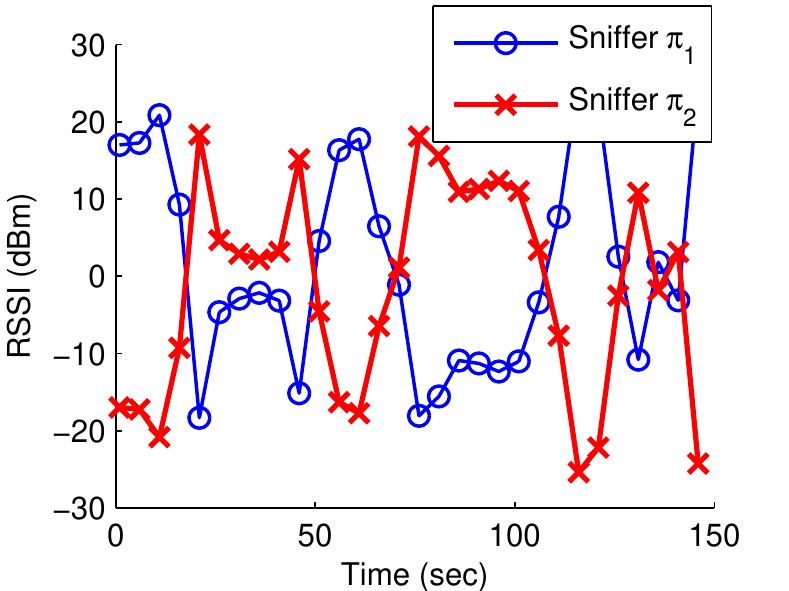} \\
(a) & (b)  & (c) \\
\end{tabular}
\caption{RSSI measurements: (a) Raw data, (b) aggregated by 5 sec, and (c) extracted after aggregation.}
\label{Fig:Signals}
\end{figure*}

The mobility data preprocessing phase starts with adding signals from multiple sniffers on top of each other for the same time window. Fig.~\ref{Fig:Signals}-a shows an example raw RSSI data collected from $\pi_{1}, \pi_{2}$ for a short time period of 150 sec for a short distance deployment (about 10m between the sniffers). As it can be seen, the RSSIs change in a way that the sniffer with stronger signals differs by time. One way of understanding the movement from one sniffer to another could be extracting peak points to find the times where a beacon is very close to particular sniffers. However, this brings other complexities and problems caused by real life challenges. As it can be seen in Fig.~\ref{Fig:Signals}-a, raw data has much noise which can be due to many reasons including the inefficiencies of wireless transmission, antenna orientation, interference, physical obstacles (e.g., walls), instability of the carried mobile devices, and so on. Therefore, finding exact points when a person passes by a sniffer has challenges in real world data such as having multiple peak points close to each other, high variances, and sometimes even not having any peak point when someone is close to a sniffer.

To analyze the data by eliminating noise and variances in the RSSIs, the values are aggregated by a sampling time $t_{s}$. $t_{s}$ is a parameter that can be set according to the distance between the sniffers and expected walking time, such that $t_{s}$ should contain at most one movement (from $\pi_{1}$ to $\pi_{2}$ or vice versa). In addition to eliminating noise and variances, data sampling and aggregation provide efficiency in the computation and discretization of the measurements for enabling the latter phases of the approach. Considering $t_{s} = 5$ sec as an example, Fig.~\ref{Fig:Signals}-b shows the aggregated signals. In this figure, the times when RSSI of one $\pi$ is higher are more visible. To visualize the differences between the RSSIs more clearly, the two signals are extracted from each other as shown in Fig.~\ref{Fig:Signals}-c. Sniffer $\pi_{1}$ represents the values of $\pi_{1}$ after $\pi_{2}$ is extracted from it. This provides a more visible and symmetric view on the signal strength differences in various times. For instance, in $T=[0,10]$ sec, RSSI of $\pi_{1}$ is higher than the RSSI of $\pi_{2}$. As RSSI values grow higher when the person with the mobile device is closer, we can infer that the person is closer to $\pi_{1}$ compared to $\pi_{2}$ at that time period.

\subsection{Wireless Fingerprinting}

As the multiple sniffers deployed are identical to each other, the RSSIs give us an insight on to which $\pi$ a person is closer to. The wireless fingerprinting is based on the fact that RSSIs are inversely proportional with distance from person $P$ to the sniffer $\pi$. For each time interval $T$ ($|T| = t_{s}$), wireless sniffer fingerprint of the person $P$ consists of the list of sniffers $S^{P}_{T} = \{\pi_{1}, \pi_{2}, ..\}$, where the sniffers are listed in descending RSSI order such that the sniffer with the highest RSSI is placed at the beginning of the list, while the one with lowest RSSI is placed at the end. A sniffer's appearance in this list means that the sniffer received signal(s) from the wireless device of $P$ during the time interval $T$. As it can be seen in Fig.~\ref{Fig:Method}, the number of sniffers at each time interval may or may not be the same. Furthermore, the sniffer with higher RSSIs may alter as this case is shown in the third time interval. The outcome of this phase is time series data where each time interval includes a list of sniffers as the wireless fingerprints. The wireless fingerprints are created for every person that is observed by the system.

\subsection{Movement Detection and Space Correlation}

Using the wireless sniffer fingerprints created in the previous phase, we define the status of person $P$ at any time interval $T$ as follows.

%
%

\[
Status(P,T) =
   \begin{cases}
     Static &\text{if $f(S^P_T, k) = f(S^P_{T_{prev}},k)$};\\
     Dynamic &\text{otherwise,}
   \end{cases}
\]
where $1\leq k \leq n$ and $n$ is the number of sniffers. The function $f(S^{P}_{T}, k)$ gives the first $k$ element of the list $S^{P}_{T}$. In our approach, we assume $k=1$ based on the observation that the sniffers that are not close to a person has either fewer RSSIs or no RSSI with some randomness. On the other hand, the closest sniffer ($\pi$ with on average higher RSSIs) is the most reliable source of input as this sniffer can receive more wireless packets. Moreover, one can simply suggest that if $\pi$ with highest RSSI changes for $P$, this is possibly because of a movement which makes $P$ closer to a different $\pi$. Hence, setting $k=1$ means that if the $\pi$ with the highest RSSI stays the same, $P$ is static.

Now, let us define the space correlation between multiple dynamic people. Considering the above assumption, dynamic means that $f(S^P_T, k) \neq f(S^P_{T_{prev}},k)$. We define a space correlation between two people $P_i$ and $P_j$ as follows.

\[
C(P_i,P_j) =
   \begin{cases}
     true & \text{if $\Big(f(S^{P_i}_{T},k) = f(S^{P_j}_{T},k)\Big) \wedge$} \\
         & \text{$\Big(f(S^{P_i}_{T_{prev}},k) = f(S^{P_j}_{T_{prev}},k)\Big)$};\\
     false &\text{otherwise,}
   \end{cases}
\]
where $1\leq k \leq n$ and $n$ is the number of sniffers. Similar to the status definition, we assume $k=1$ considering the correlation based on the closest sniffer for the pair $(P_i,P_j)$.


For each pair $(P_i, P_j)$ for any time interval $T$, if $Status(P_i,T) = Status(P_j,T)$ and $C(P_i,P_j) = true$, the pair is considered to have a movement together. The computation is iterated for all pairs so that the movement group sets $G= \{P_1, P_2, ...\}$ are created for any time interval. For detected movements with the set $G$, $|G|=1$ for alone movements and $|G|>1$ for together movements. In summary, the approach detects group mobility as well as individual movements based on the aforementioned phases of data preprocessing, wireless fingerprinting, and using the movement status and space correlation information.

\section{Experimental Study}
\label{sec:Experiments}

\begin{figure}
\centering
\includegraphics[width=0.5\textwidth]{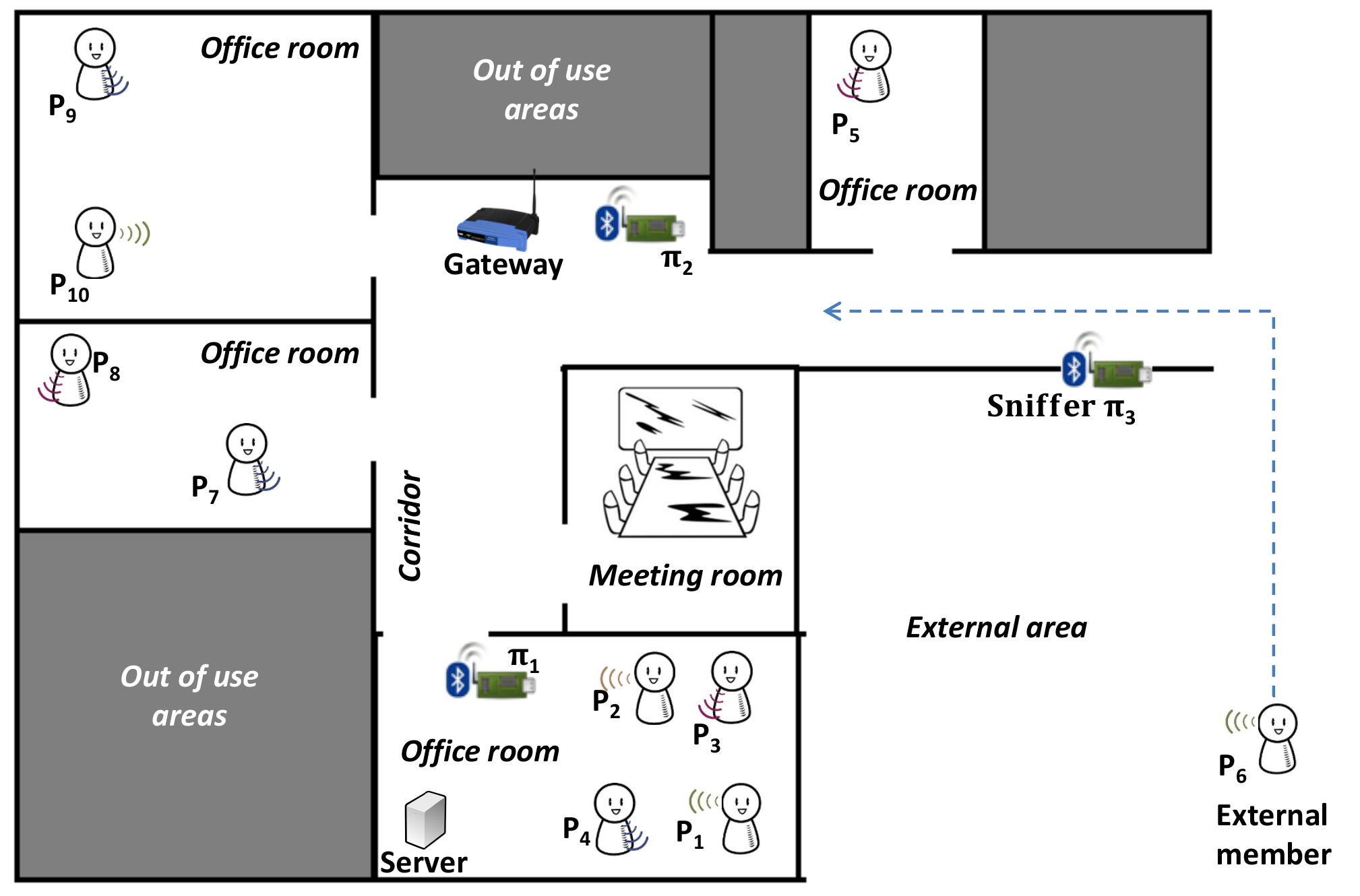}
\caption{The experimental setup in the office environment: 3 sniffers, 10 people (from 4 rooms \& 1 external).}
\label{Fig:ExpSetup}
\end{figure}

\subsection{Experimental Setup}
\label{subsec:ExperimentalSetup}
For the experiments we implement a packet sniffing program in Python for Raspberry Pi platforms. We use BLE sensing technology as an example for the prototype of our system. The software consists of 3 components: {\em packet monitor}, {\em packet filter and decoder} and {\em data reporter}. The packet monitor captures all types of BLE packets. The packet filter and decoder parses only BLE advertising packets and drops other types of BLE packets. The data reporter sends newly arrived packets to the crowd mobility database (CouchDB) through the gateway for performing data analytics. The designed software components are running as background processes on Raspberry Pi version 3 which has a built-in BLE module.

After setting up the system and preliminary analysis, we conduct two types of experiments: (a) controlled experiment for collecting ground truth and (b) real-world experiment to collect data from people for a longer period of time. In the controlled experiment 10 beacon nodes are carried together (all put in a small box) and one person performs 10 and 20 sec walks and 5 sec runs (in 2.5 minutes time in total) between 2 sniffers which are placed about 10 m away from each other. Our purpose is to understand if the results for the beacons are similar and consistent with each other. While they are different beacon nodes, they are at any time placed next to each other, therefore the expectation is to have similar results. Later, we conduct a real-world experiment in the office environment, where we collect Bluetooth data from 10 participants who carry beacons for a two-week period (during work hours). The two participants carry 2 beacons at the same time to help us understand how accurate the results are. We want to know how frequent the group movements occur as opposed to individual movements in the office environment. Moreover, we aim to see the reflection of the setting of the office such as people who stay in the same room or people who stay alone as well as people who mostly walk together (e.g., going to meetings together).

Fig.~\ref{Fig:ExpSetup} shows the experimental setup and the setting of the office environment. 3 sniffers are located in the corridors where people mostly walk without interruptions. The gateway is placed such that each $\pi$ can transmit their arrived packets and later the data is forwarded to the backed server (in the office room) through the gateway. Rooms have different sizes (4 people, 2 people, single room) and one participant is called as an {\em external member} since the office room of the participant is out of the range. However, the external member walks through these corridors from time to time due to working in the same group and also for going to lunch. Lastly, there exist a meeting room on the middle which can be used by any participants. The experimental setup (e.g., locations of the sniffers) is static in controlled and real-world experiments for this initial evaluation of the feasibility of our approach and simple comparison of the results from the different experiments. On the other hand, we believe that the approach can be easily applied to various indoor and outdoor setups including different types of conditions and obstacles.

The default parameters in our experiments are as follows. The distance between $\pi_1$ and $\pi_2$ is 10.5 m and between $\pi_2$ and $\pi_3$ is 9.5 m. $t_{s}$ is empirically set as 20 sec. The advertising interval of the beacons are set as 100 msec and the transmission power of the beacons is set as 4 dBm. While the approach does not rely on high transmit power or very short advertising interval, we observe that for weak transmission powers packet arrival rate may decrease even when the person is closer to the $\pi$ (e.g., 2 m from the $\pi$). For the controlled experiment which takes 2.5 min, we observe 40\% packet arrival ($\sim1$ MB JSON file) for weak transmission power of -12 dBm compared to strong transmission power of 4 dBm ($\sim2.5$ MB JSON file). Experimental results do not involve any processes for filtering out or averaging results from multiple runs.

\begin{figure*}
\centering
\begin{tabular}{ccc}
\includegraphics[width=0.33\linewidth]{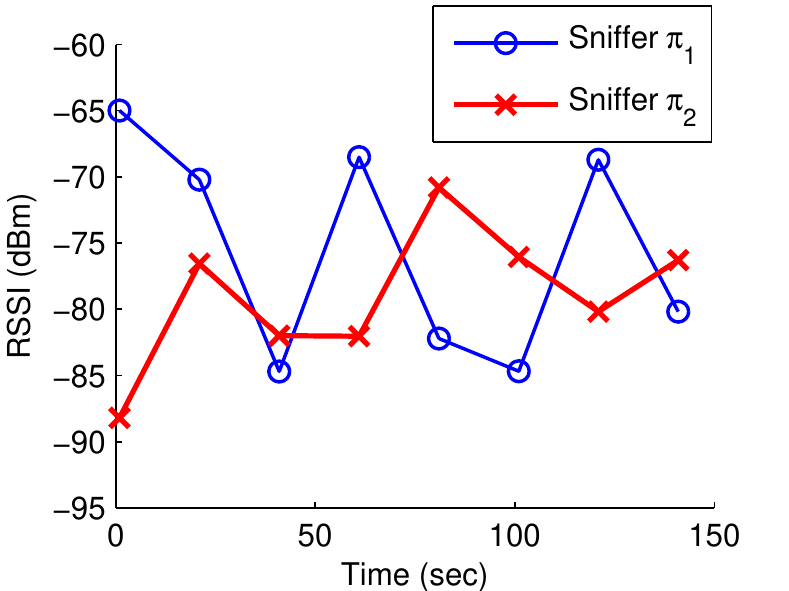} &
\includegraphics[width=0.33\linewidth]{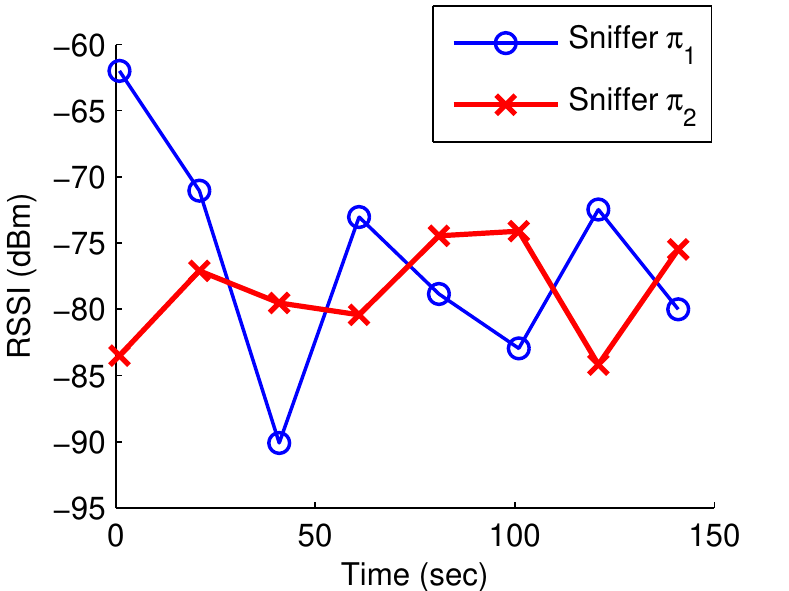} &
\includegraphics[width=0.33\linewidth]{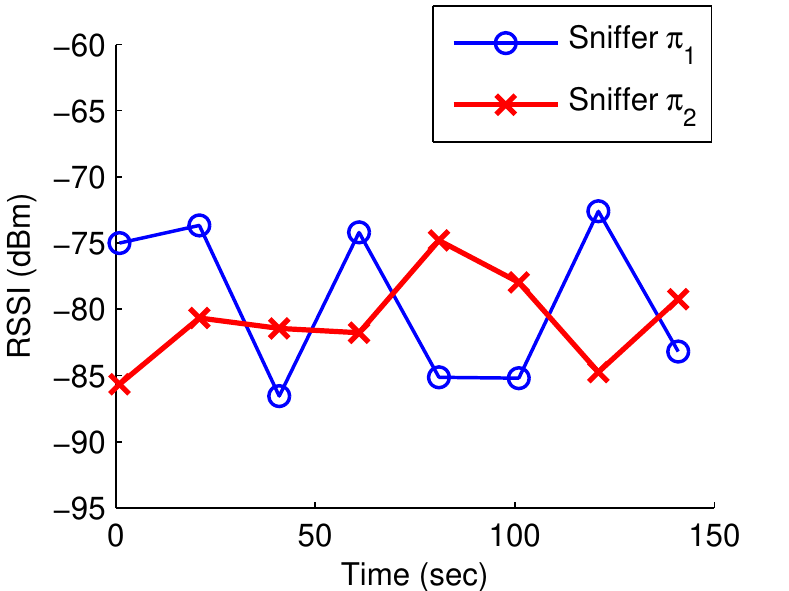} \\
(a) & (b)  & (c) \\
\end{tabular}
\caption{Signals of 3 beacons after aggregation (20 sec): (a) Beacon 1, (b) Beacon 4, and (c) Beacon 9.}
\label{Fig:SameSignals}
\end{figure*}

\subsection{Metrics}
We define the first metric for analyzing the results based on the space correlation of wireless fingerprints. {\em Similarity score} is defined for the set of wireless fingerprints $S_T^{P_i}, S_T^{P_j}$ of a pair $(P_i, P_j)$ (when both have fingerprint for a particular time interval $T$) for any duration $\lambda=\{T_1, T_2,..,T_n\}$ ($\lambda$ can be a short or a long duration). The maximum possible points is defined as the case when they have both the exact same fingerprints (e.g., $S_{T}^{P_i}=\{\pi_2,\pi_1,\pi_3\}, S_{T}^{P_j}=\{\pi_2,\pi_1,\pi_3\}$ for all time intervals in $\lambda$ except when both $P_i$ and $P_j$ have no measurements. The matching reward depends on the precedence such that the first match (e.g., both fingerprints start with $\pi_2$) is rewarded with 7 points, second match (e.g., both fingerprints has second element $\pi_1$) is rewarded with 2 points, and the third match is rewarded with 1 point. Note that the third match reward is not used for the controlled experiment as there are only 2 sniffers in this experiment. Moreover, for a particular time interval $T$, a match cannot occur after a mismatch. For instance, if the first sniffers in the two fingerprints do not match, the rest is labeled as mismatches due to the precedence of the first match over the others. While this scheme can be modified for different experiments, in particular for the controlled experiment it gives a valuable insight on the reliability and consistency of the fingerprints collected from the beacons.

We analyze the {\em number of movements} (our second metric) detected based on the group sizes such as individual movement, 2 people movements, 3 people movements, and so on. To understand if the group movements reflect the social setting in the office environment, we define the metric {\em movement intersections}. The metric movement intersections ($MI$) is defined as follows.

\[
MI(P_i,P_j) = \frac{\|M(P_i) \cap M(P_j)\|}{\|M(P_i)\|},
\]
where $M(P_i)$ is the set of all detected movements of $P_i$ and $M(P_i) \cap M(P_j)$ is the set of common movements of $P_i$ and $P_j$.

The above metric takes alone walks into account, such that even when only $P_i$ walks alone ($|G|=1$), $MI$ decreases (since the divisor $M(P_i)$ is the set of all movements of $P_i$). In order to observe pairs in the group movements, we define another metric called {\em together movement intersections} ($TMI$), where together movement means the group movements of the size at least 2, as follows.

\[
TMI(P_i,P_j) = \frac{\|M(P_i) \cap M(P_j)\|}{\|TM(P_i)\|},
\]
where $TM(P_i)$ represents the together movements of $P_i$. As it can be seen, similarity score produces symmetric results for the pairs (similarity score of $(P_i,P_j)$ is equal to $(P_j,P_i)$), while $MI$ and $TMI$ depends on the perspective of $P_i$ or $P_j$ as the divisor changes.

\subsection{Performance results}

\begin{figure}
\centering
\includegraphics[width=0.5\textwidth]{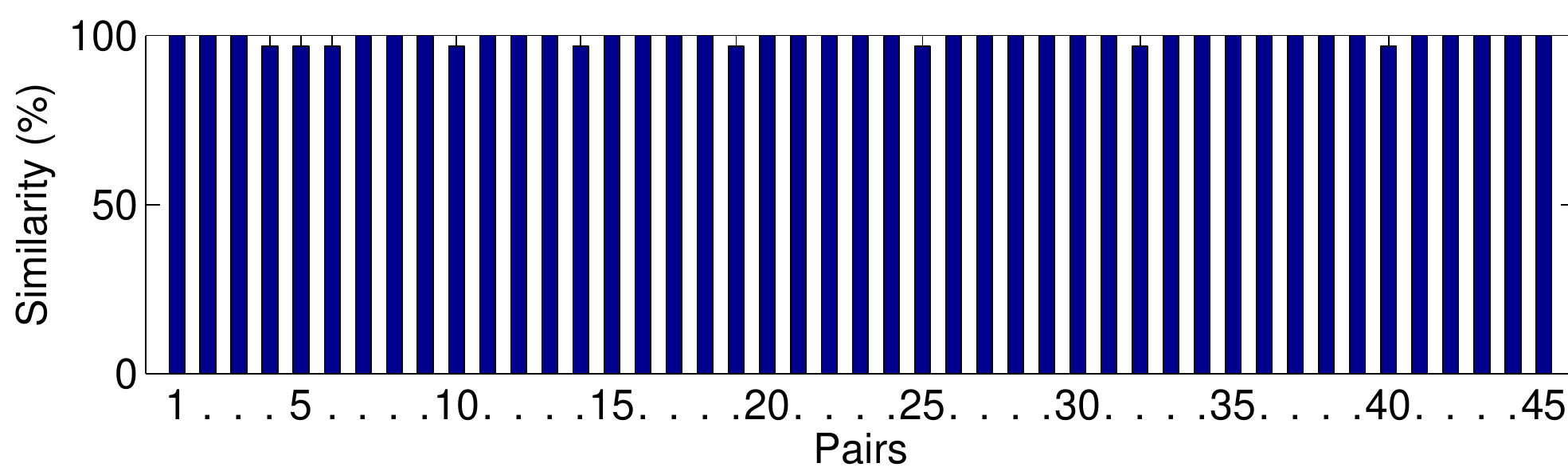}
\caption{The 10 beacon experiment: group movement similarity scores of all pairs of beacons.}
\label{SimilarityPairs}
\end{figure}

\subsubsection{Experiment 1: 10 beacons together}

\begin{figure*}
\centering
\begin{tabular}{ccc}
\includegraphics[width=0.33\linewidth]{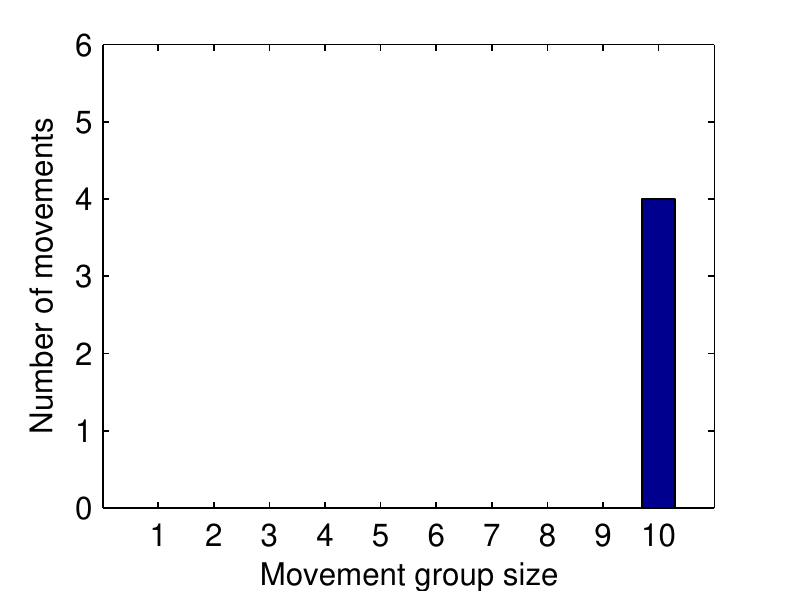} &
\includegraphics[width=0.33\linewidth]{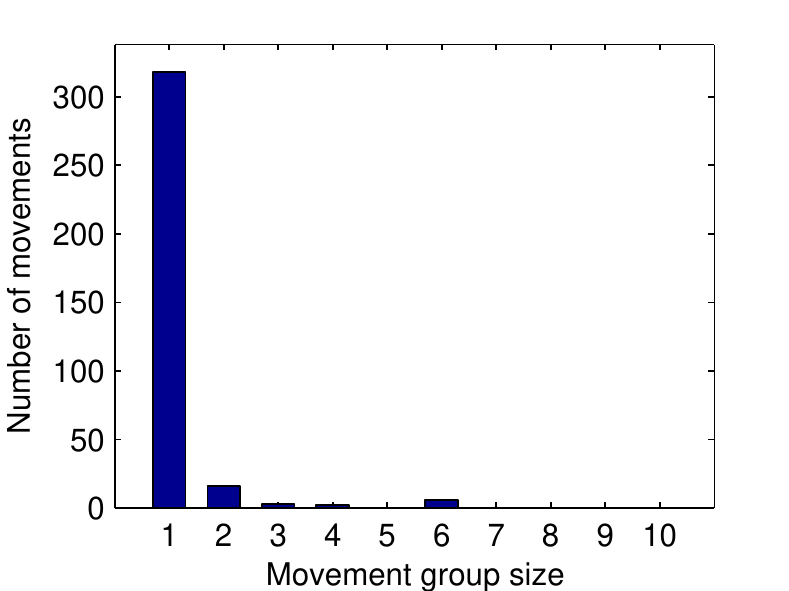} &
\includegraphics[width=0.33\linewidth]{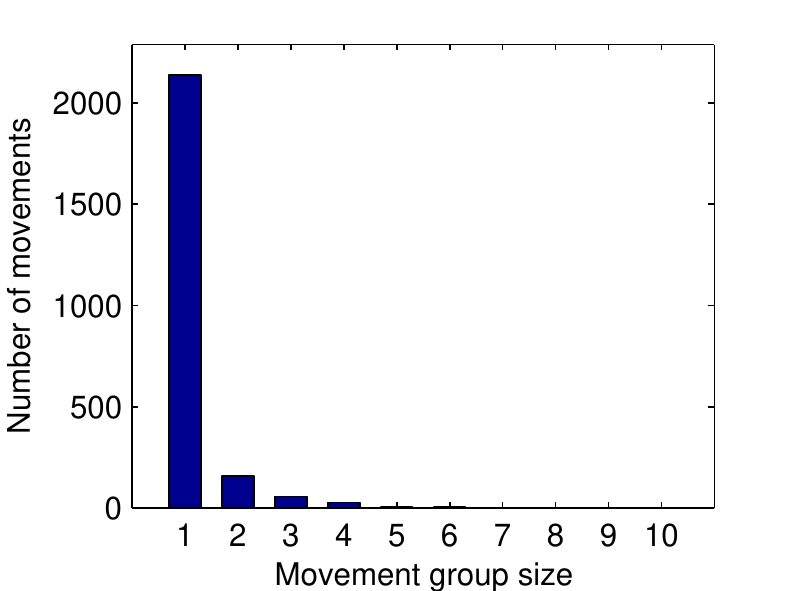} \\
(a) & (b)  & (c) \\
\end{tabular}
\caption{Movements detected for (a) 10 beacons together, (b) 1 day measurement  (c) 2 weeks measurement (10 people).}
\label{Fig:GroupMovementsBar}
\end{figure*}

We compare the RSSI measurements collected from 10 beacons that are placed all together while moving between 2 sniffers. After walking back and forth between the sniffers, we stop data collection process and visualize the signals from both sniffers by placing them on top of each other. The raw data we observe have similar patterns from measurements of different beacons such that the peak points, where the signals are much stronger, are placed in similar time frames. These patterns becomes more clear after the sampling and aggregation phase. Fig.~\ref{Fig:SameSignals} shows the signal patterns after sampling and aggregation for 3 randomly picked beacons (beacon $B_1, B_4,$ and $B_9$) out of the 10 beacons. We observe for each time frame which sniffer received stronger signals (e.g., $T=[0,20]$ is shown at $0^{th}$ sec, $T=[20,40]$ is shown at $20^{th}$ sec, and so on). As you can see in Fig.~\ref{Fig:SameSignals}, while the overall signal strengths differ from one beacon to another, for every time frame, the sniffer with stronger signal is the same for all 3 beacons. This shows that for this scenario wireless fingerprinting of the 3 beacons results in the exact same set of fingerprints. Similarly, dynamic/static status of the beacons are the same and they have high space correlation as these phases are products of the wireless fingerprints. The resulting set of wireless fingerprints is as follows.
\begin{align*}
   WF(B_1)=& \Big\{ \{\pi_1,\pi_2\} , \{\pi_1 , \pi_2\} ,\{\pi_2, \pi_1\} , \{\pi_1, \pi_2\} , \{\pi_2, \pi_1\}, \\
        &\{\pi_2, \pi_1\} , \{\pi_1, \pi_2\} , \{\pi_2, \pi_1\} \Big\},
\end{align*}
where $WF(B_1)$ is the wireless fingerprints of $B_1$. Note that $WF(B_1)=WF(B_4)=WF(B_9)$.

We expect the produced wireless fingerprints to be similar to each other since the beacons are carried together. If the produced signals does not have the similarity, than one can say that the approach cannot produce consistent and reliable results. On the other hand, if the wireless fingerprints are the same (during these multiple walks in the controlled experiment), this proves that they also produce the same movement detection and space correlations (as they are solely based on fingerprints) and therefore consistent group mobility results even when using different mobile devices. To see if the produced wireless fingerprints are similar to each other, we compare all possible pairs of beacons (45 pairs) based on the similarity score. Fig.~\ref{SimilarityPairs} shows the similarity scores of the 45 pairs. We observe almost 100\% similarities for every pair, meaning that the wireless fingerprints are almost exactly the same and the 10 beacons are marked as together during all time intervals as they are expected to be. The average similarity score (average of all pairs) is 99.4\% for $t_{s}=20$ sec, 89.5\% for $t_{s}=10$ sec, 83.8\% for $t_{s}=5$ sec, 78.7\% for $t_{s}=2$ sec, and 71.1\% for $t_{s}=1$ sec. We proceed with using 20 sec as the aggregation time since it provides the most reliable results with an expected level of granularity.

%

%

\subsubsection{Experiment 2: Group mobility detection}

Based on our proposed approach, we detect the group movements in the controlled experiment as well as the real-world experiment. The goal is to successfully detect the people walking together or alone. As shown in Fig.~\ref{Fig:GroupMovementsBar}-a, 4 different movements (out of 6) are detected. The system missed capturing the back and forth running in 5 sec. This shows the trade-off between capturing all the movements vs. reliable and consistent capturing of expected movements. We also observe that (as shown in Fig.~\ref{Fig:GroupMovementsBar}-a) the system does not only capture the movements, but also successfully mark all beacons as together. The group size of each of the 4 movements is equal to 10.  Hence, we observe that the approach successfully captures group mobility of the 10 beacons.

Fig.~\ref{Fig:GroupMovementsBar}-b shows the movements detected in one day real-world experiment. There exist more than 300 alone walks while around 20 movements of 2 people together. Fig.~\ref{Fig:GroupMovementsBar}-b shows the movement detections for the two-week period. This figure demonstrates more significant outcomes in terms of the difference between alone walks and together walks with group sizes of 2, 3, 4, 5, and 6. Moreover, there exist no group mobility behavior for more than 6 people, even though it is theoretically possible as the 10 participants work in the same environment.

\begin{figure}
\centering
\includegraphics[width=0.45\textwidth]{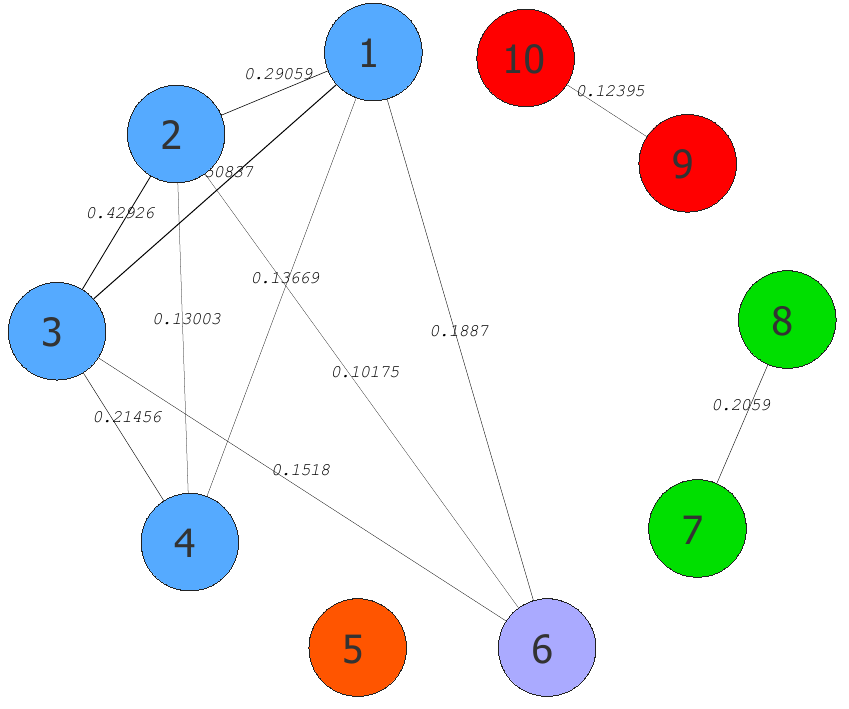}
\caption{Movement intersections (MIs: including alone walks) for 10 people in 2 weeks (values in the range [0,1]).}
\label{Fig:MovementIntersection}
\end{figure}

\begin{figure}
\centering
\includegraphics[width=0.5\textwidth]{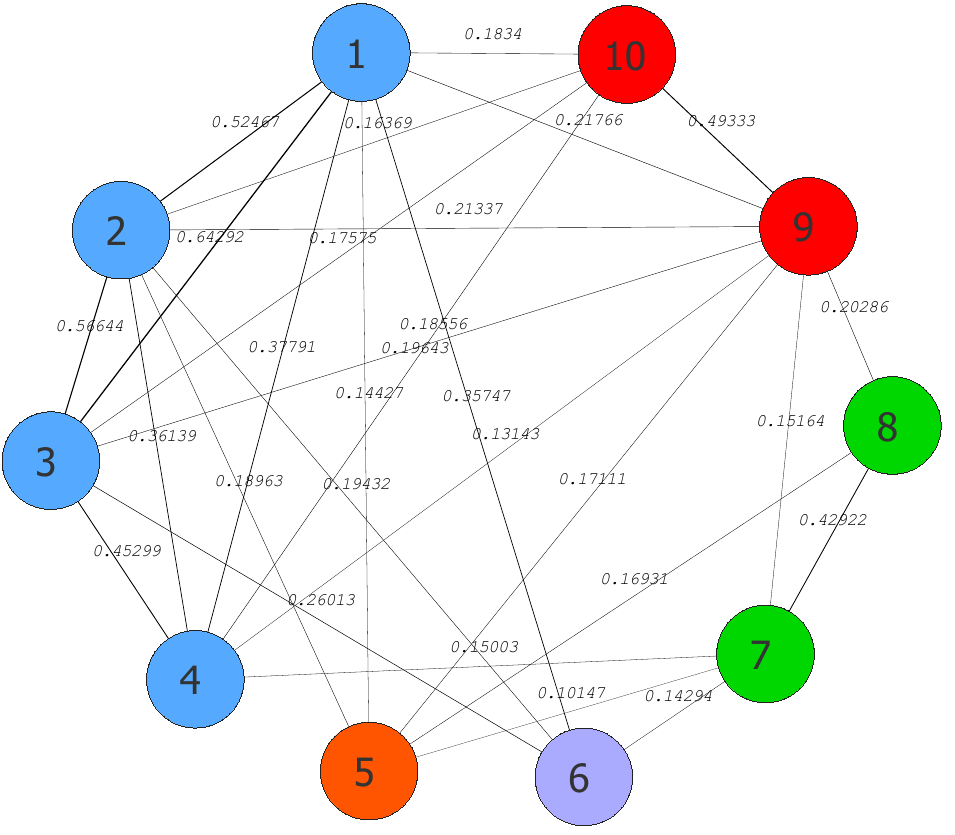}
\caption{Together movement intersections (TMIs: excluding alone walks) for 10 people in 2 weeks (values in the range [0,1]).}
\label{Fig:TogetherMovementIntersection}
\end{figure}

To see the reflection of the experimental setting described in Sec.~\ref{subsec:ExperimentalSetup} (Fig.~\ref{Fig:ExpSetup}) on the detected group mobility, we analyze the results of the two-week period based on the group movement intersection metrics ($MI$, $TMI$)  and visualize the results in social networks. Note that the measurements using these metrics are pairwise relative. For instance, between 2 different pairs, one pair having a very high $MI$ compared to the other does not show that the pair with the higher $MI$ has more group movements, but it shows that when the pair walks they mostly prefer to walk together. For the visualization purpose, the intersections between pairs which are less than 0.1 are omitted. Furthermore, the resulting graph is actually a directed graph where each pair has two edges with different weights. We took the average of these two edge weights and visualized the network as a nondirected graph for simplicity. Fig.~\ref{Fig:MovementIntersection} shows the social network created based on $MI$s. In this figure nodes represent people (e.g., 2 represents $P_2$), edges represent pairs that have more than 0.1 $MI$, and edge weights represent $MI$s. The nodes representing the people who are working in the same room are marked with the same color (5 colors in total). We observe that the people who work in the same room tend to walk together with each other as opposed to walking with people from other rooms. In particular, the group movement tendency is clearly seen for $P_1, P_2, P_3, P4$ and the highest $MI$ is observed between $P_2$ and $P_3$, followed by $P_2$ and $P_1$. There is only one exception to the working rooms, which is the pairwise relations of the external member $P_6$. While $P_6$ does not work in the same room, he/she has a tendency to walk together with $P_1$, $P_2$, and $P_3$ (e.g., going for lunch together).

Fig.~\ref{Fig:TogetherMovementIntersection} illustrates the social network created based on the $TMI$s which exclude the alone walks from calculations. In this figure edges represent pairs that have more than 0.1 $TMI$s, and edge weights represent $TMI$s. The outcome is a more complete graph of group movement relations. However, by looking at the $TMI$ values  between pairs specifically, we observe the similar pattern of group movement tendencies such that the people who are located in the same office rooms have more group movements together. The highest interaction stays between $P_2$ and $P_3$, followed by $P_2$ and $P_1$. Another observation is related to the alone movement tendencies. For instance, the difference of $MI$ and $TMI$ results for $P_9$ and $P_{10}$ are very significant since $P_9$ and $P_{10}$ mostly prefer to walk alone. As the alone movements are excluded, their $TMI$ value goes up to 0.49.

In order to analyze the reliability of the results for the real-world experiments, we ask $P_1$ and $P_2$ to carry 2 beacons at the same time throughout the experiment. For each person, the 2 beacons have different types, one is larger, and one of them are older, while the other one is new with full battery. We analyze the beacons' similarity scores as well as $TMI$s. In the ideal case for each person, both the similarity scores and the $TMI$ values should be exactly the same (100\% accuracy). The resulting similarity scores are 94.80\% for $P_1$'s beacons, 95.97\% for $P_2$'s beacons, and the average accuracy is 95.39\%. The results for $TMI$s are 0.78 for $P_1$, 0.80 for $P_2$, with the average accuracy of 79\%.

\section{Conclusion}
\label{sec:Conclusion}

This paper proposes a new approach to detect group mobility using multiple wireless sniffers. We propose a method which creates wireless fingerprints from the collected mobility data and captures group movements of people as well as their individual walks. We implement a prototype system which collects human mobility data from BLE beacons. The performance evaluation based on the controlled and real world experiments shows that the proposed approach determines the individual movements as well as the group mobility characteristics with accuracies of 95\% for space correlation and 79\% for group mobility.

\section{Acknowledgment}
\parpic{\includegraphics[width=0.23\linewidth,clip,keepaspectratio]{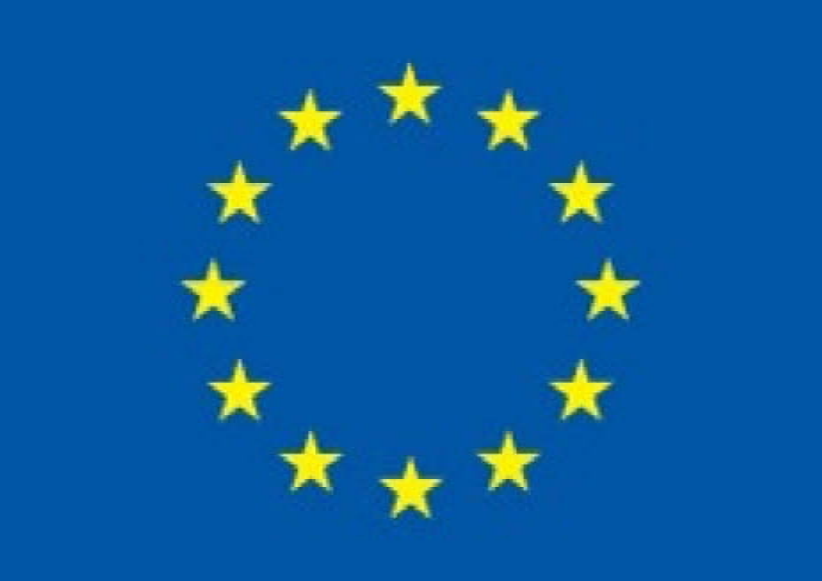}}
\noindent This work has received funding from the European Union's Horizon 2020 research and innovation programme within the project ``Worldwide Interoperability for SEmantics IoT" under grant agreement Number 723156. 

\bibliographystyle{IEEEtran}



\end{document}